\documentclass[12pt]{article}

\usepackage{color,epsfig}
\usepackage{graphicx}

\definecolor{violet}{rgb}{0.4,0,0.6}
\definecolor{vert}{rgb}{0,0.4,0.0}
\definecolor{navy}{rgb}{0.0,0.0,0.4}

\def\colbrun#1{\textcolor[named]{Brown}{#1}}

\def\be{\begin{equation} }

\def\fe{\end{equation}}

\def\spose#1{\hbox to 0pt{#1\hss}}\def\lta{\mathrel{\spose{\lower 3pt\hbox
{$\mathchar"218$}}\raise 2.0pt\hbox{$\mathchar"13C$}}}  \def\gta{\mathrel
{\spose{\lower 3pt\hbox{$\mathchar"218$}}\raise 2.0pt\hbox{$\mathchar"13E$}}}

\def\mm{{\color{red}m}} \def\ee{{\color{red}e}}
\def\aalpha{{\color{red}\alpha}}
\def\gg{{\color{red}g}}
\def\tt{{\color{blue}t}} \def\ttau{{\color{blue}\tau}}
\def\RR{{\color{vert}R}}  \def\lel{{\color{vert}\ell}} 
\def\aa{{\color{vert}a}}    
\def\vv{{\colbrun{v}}} 
\def\nn{{\colbrun{n}}}
\def\NN{{\color{violet}N}} \def\Ttheta{{\color{violet}\Theta}} 
\def\MM{{\color{violet}M}} \def\rrho{{\color{violet}\rho}}
 \def\EE{{\color{violet}E}} 
\def\LL{{\color{red}L}}
\def\varepsilo{{\color{red}\varepsilon}}

\begin{document}

\begin{center}\colbrun{\em Contribution to 
{\bf Time in Science, Anthropology, Religion, Arts},\\
ed. {\rm A. Nicolaidis}, SR21 workshop, Thessaloniki and Athens. }
\\[0.8cm]
 \textcolor{red}{\Large Objective and subjective time\\[0.3cm]
in anthropic reasoning.
}    \\[0.8cm]
 \underline{Brandon Carter} \\[0.5cm]
 \textcolor[named]{ForestGreen}{LuTh, Observatoire Paris - Meudon }
  \\[0.8cm]
 \colbrun{\em September, 2007. }  \\

\end{center}

{\bf Abstract.}
The original formulation of the (weak) anthropic principle was 
prompted by a question about objective time at a macroscopic level, 
namely the age of the universe when ``anthropic'' observers such as 
ourselves would be most likely to emerge. Theoretical interpretation
of what one observes requires the theory to indicate what is expected,
which will commonly depend on where, and particularly when, the
observation can be expected to occur. In response to the question of 
where and when,  the original version of the anthropic principle
proposed an {it a priori}  probability weighting proportional to the 
number of ``anthropic'' observers present. The present discussion 
takes up the question of the time unit characterising the biological
clock controlling our subjective internal time, using a revised
alternative to a line of argument due to Press, who postulated
that animal size is limited by the brittleness of bone.
On the basis  of a static support condition depending on the tensile
strength of flesh rather than bone, it is reasoned here that our size
should be subject to a limit inversely proportional
to the surface gravitation field $\gg$, which is itself found 
to be proportional (with a factor given by the 5/2 power of the
fine structure constant) to the gravitational coupling constant. 
This provides an animal size limit that will in all cases be of the
order of a thousandth of the maximum mountain height, which
will itself be of the order of a thousandth of the planetary radius.
The upshot, via the (strong) anthropic principle, is that the need
for brains, and therefore planets, that are large in terms
of baryon number may be what explains the weakness of gravity
relative to electromagnetism.

\vfill\eject

\section{Introduction: fundamental parameters.}

                         In this interdisciplinary forum for the 
discussion of various kinds of time, one of the first questions 
that comes to mind is that of the relationship between external 
time of the kind measured by physicists in the objective world, 
and the internal psychological time characterising successive 
instants of conscious perception by a human (or other comparable) 
mind. As this is something I had previously thought about in the 
context \cite{Carter05} of anthropic reasoning for astronomical 
and other purposes (notably in the development 
\cite{Carter04,Carter06} of a microanthropic principle such as 
seems necessary for the interpretation of quantum theory) my 
objective here will be to offer a brief account of what emerges
from that point of view.

Before recapitulating what is meant by the notion of anthropic
reasoning, I would start by explaining that this essay will
will take for granted the usual paradigm whereby it is assumed
that the mind, as perceived from inside, corresponds, in a 
material physical world, to an organism of the kind known as a 
brain. More specifically, the motivation for the following 
discussion derives from the supposition (invoked in previous 
relevant work by Dyson \cite{Dyson79} and Page \cite{Page96}) 
that mental processes are describable in terms of instants of 
perception characterised by a finite duration, 
$ \widetilde\ttau=\delta \tt\, ,\label{delt} $
of time $\tt$ -- interpretable as a basic biological clock unit --
as measured in the physical world wherein the brain is situated. 
In the human case it would seem that the shortest biological 
clock timescale on which coherent mental processes occur is of 
the order of a fraction of a second, which is of course why the 
latter has been chosen as the standard time measurement unit

It has for more than a century been possible, and for more that
half a century been common usage, to take the fundamental
physical measurement units of mass, of length, and 
(for our present purpose most pertinent) time to be those 
of what is known as the Planck system, which attributes unit
value to three particularly important fundamental parameters.
The first of these is the coupling constant $G$ that was crucial
for what was historically the earliest satisfactory physical theory
namely Newtonian gravitation. The second is the propagation
speed $c$ that was crucial for the next great physical theory,
namely that of Maxwellian electromagnetism, which was subsequently
unified with gravitation by Einstein. The third is the more 
mysterious parameter designated $\hbar$, of which the importance
was first recognised by Planck, but of which the significance
was not properly understood until (synthesising contributions by
by Schrodinger, Heisenberg, Born and others) the principles
of modern quantum theory were finally established by Dirac.

An immediate consequence of this astounding breakthrough was
to provide an explanation of the complexity of the chemical 
behaviour of the elements, as described by Mendleef's puzzling 
periodic table, in terms of just a single dimensionless coupling 
constant, namely the charge of the electron, whose square, the 
fine structure constant $\aalpha_{\rm e}=\ee^2$, is given 
approximately, in these units, by $\ee^2\doteq1/137$, while a 
complete account of low energy physics on a local scale required 
only one more dimensionless parameter namely the electron proton 
mass ratio $\mm_{\rm e}/\mm_{\rm p}\doteq 1/1830$ which is 
important for the properties of liquid helium. These values, and
those of the Yukawa coupling constant $\aalpha_{_{\rm Y}}\doteq 2/7$ 
and the pion proton mass ratio $\mm_\pi/\mm_{\rm p}\doteq 1/7$ that 
play an analogous role in the much less complete theory of strong 
interactions introduced about the same time, were fundamental in 
the sense of being obtainable only  by empirical observation. 
Early hopes that they would soon be obtainable by calculation from 
some deeper theory remain unfulfilled half a century later. 

Although there has of course been progress in the derivation of 
Yukawa's rudimentary pion coupling model from more sophisticated strong 
interaction theories involving quarks and gluons, this has been done 
only by introducing even more adjustable parameters than before, so 
the outcome is that the value of the effective coupling constant 
$\aalpha_{_{\rm Y}}$ still remains something  known only by empirical 
measurement . Therefor to account for the value of this and other such
quantities, there is now more interest than ever in what I called the
strong anthropic principle \cite{Carter74}. This means an approach 
whereby the quantities in question are postulated to have values that 
vary over observationally inaccessible parts of what recently come to be 
known \cite{Davies04} as a multiverse, within which our own  region has
 been selected by the existence of -- and therefor environmental 
suitability for -- observers like ourselves. 

As the prototype candidate for such a selection effect, I had drawn 
attention \cite{Carter67} to the observed Yukawa coupling relation  
{\be \aalpha_{_{\rm Y}}\doteq 2\mm_\pi/\mm_{\rm p}\, , \label{nucmarge}\fe}
that is well known as the condition for the nuclear coupling to be 
marginally strong enough for the formation of a deuteron by binding of 
a proton and a neutron within the distance fixed by the pion mass.
As consequently remarked by Dyson \cite{Dyson71} and confirmed by 
subsequent calculations \cite {PPBR91}, a  relatively small
respective decrease or increase in this coupling would  suffice
to provide a chemically sterile universe consisting exclusively of 
hydrogen, or containing none at all. 

Whereas most such fundamental coupling coupling constants
were found to have values  more or less comparable with 
unity, Dirac was impressed by the fact that there is 
an exception. Compared with their electromagnetic 
attraction, the gravitational attraction between an electron
and a proton is weaker by an enormous factor of the order 
of $10^{40}$, a number that is interpretable as roughly the
inverse of the product of the mass of these particles, as given
in Planck units roughly by $\mm_{\rm e}\approx 10^{-22}$
and  $\mm_{\rm p}\approx 10^{-19}$, so that their geometric
mean square value is given in very round figures by
{\be  \mm_{\rm e}\, \mm_{\rm p}\approx  \langle \mm\rangle^2
\approx 10^{-40}\, .\label{meanm}\fe}
The point to which I wish to draw attention here -- and for which 
at the end I shall offer a tentative explanation -- is
that in terms of such Planck units the fraction of a second time 
unit characterising our mental (and other biological)
processes has a value of about the same enormous
order of magnitude, namely
{\be  \widetilde\ttau\approx 10^{40}\, .\label{forty}\fe}

\section{Dirac's coincidence}

Dirac himself drew attention, not to the coincidence I have
just mentioned,, but to another coincidence involving the same 
number, which is that it is roughly the square root of the number 
$\NN\approx 10^{80}$ of protons (or equivalently, by charge 
neutrality, of electrons) in the visible universe, meaning the volume 
characterised by the cosmological Hubble radius, which was first 
measured (though not very accurately) about the same time (three 
quarters of a century ago) that the principles of modern quantum 
theory were first established. It was Dirac's mistaken explanation 
for this highly significant coincidence that prompted me to provide 
an explicit formulation of what I called the anthropic principle.
  
Dirac's idea  was based on the supposition that we are observing at 
a random -- so presumably typical -- instant in the history of our 
expanding universe, so if (for whatever reason) the inverse of the 
gravitational coupling constant is presently equal to the square root 
of the number of particles in the visible volume determined by the 
Hubble expansion rate then it should be expected to remain  so. 
Since the visible volume will include progressively more and more 
particles as the universe expands, it would follow that the 
gravitational coupling  should become correspondingly weaker. 

When I first read about this (in Bondi's classic textbook 
\cite{Bondi} on cosmology) I realised that there was a weak link 
in Dirac's reasoning, namely his implicit adoption of what I 
would criticise as an ubiquity principle rather than the anthropic 
principle  that seems appropriate. The essential content of
what I called the anthropic principle is that, a priori,
 our location in space or time should not be considered as
random with just with respect to the corresponding ordinary
physical measure of space or time (as in what I would call an 
ubiquity principle) but with respect to
a measure that is anthropically weighted in the sense of being
proportional to the population density of individuals comparable
with ourselves. 
  
It has since become observationally clear that the conclusion to
which Dirac was drawn by his misguided line of reasoning was
indeed wrong, since his predicted (cosmological rapid) weakening of 
gravitation does not actually occur. Already, even before this was as 
obvious as it is now, the weak point in Dirac's line of reasoning 
had been noticed by Dicke \cite{Dicke61} (an advocate of another 
theory of progressive weakening of gravity, but at a much slower 
and so less easily measurable rate) who preceded my own contribution 
\cite{Carter74} in pointing out that observers comparable to ourselves
(which is what I meant by the adjective anthropic) could not
possibly have existed when the age of our expanding universe was much 
less than the lifetime of a typical hydrogen burning star, and could 
be expected to become relatively rare when the universe is much older 
that that. The (then recently discovered) reason for the lower cut off
is that such hydrogen burning is the only way of fabricating the 
medium and heavy elements of which we are made. The (more obvious)
reason for the  upper time limitation on the relevant anthropic 
weighting measure is the dependence of life systems such as ours on 
energy input from a neighbouring star, and of the fact that although 
later generations of such stars will continue to be formed they will
become progressively rarer as the matter that was originally present 
is transformed is transformed into unavailable end states such as 
cold dead neutron stars and black holes. 

Although Dirac was wrong in his conclusion of weakening of 
gravitation with time, it appears that he was right in deducing that 
there is a direct connection between the the strength of gravity and 
the number of particles in the universe as presently observed. The 
coincidence he noted  is that the value, roughly $10^{80}$, of $\NN$ 
is roughly the inverse square of the value, roughly $10^{-40}$, of 
the gravitational coupling constant that is itself given very roughly 
as square of the proton mass $\mm_{\rm p}$. Dirac's coincidence is 
thus roughly expressible in Planck units simply as
{\be  \NN\approx \mm_{\rm p}^{-4} \, .\fe}
A first step in what now appears to be the correct explanation of 
this is the recognition that, as has been known since Newton's time, 
the dynamical timescale $\tt$ associated with the free self 
gravitational acceleration of a body with mass density $\rrho$ will 
(remembering our units here are such that $G=1$) be given roughly by 
$\rrho\approx 1/\tt^2$. In the cosmological case (remembering our 
units are such that $c=1$)   the relevant Hubble radius will be of 
the same order as the age $\tt$, so the order magnitude of the mass 
in the visible volume will be given roughly by $\MM\approx \rrho\, 
\tt^3\approx \tt$ which means that the corresponding number $\NN$ 
of particles with mass $\mm_{\rm p}$ will be given by 
$\NN\approx \tt/ \mm_{\rm p}$. Dirac's coincidence is therefore 
equivalent to the (directly verifiable) observation that the present 
age, $\tt_\star$ say,  of the universe is given very roughly by
{\be  \tt_\star\approx \mm_{\rm p}^{-3}\, ,\label{unage}\fe}
which in very round figures comes out to be something like $10^{60}$.

\section{Explanation from stellar physics}

\begin{figure}
\centering
\epsfig{figure=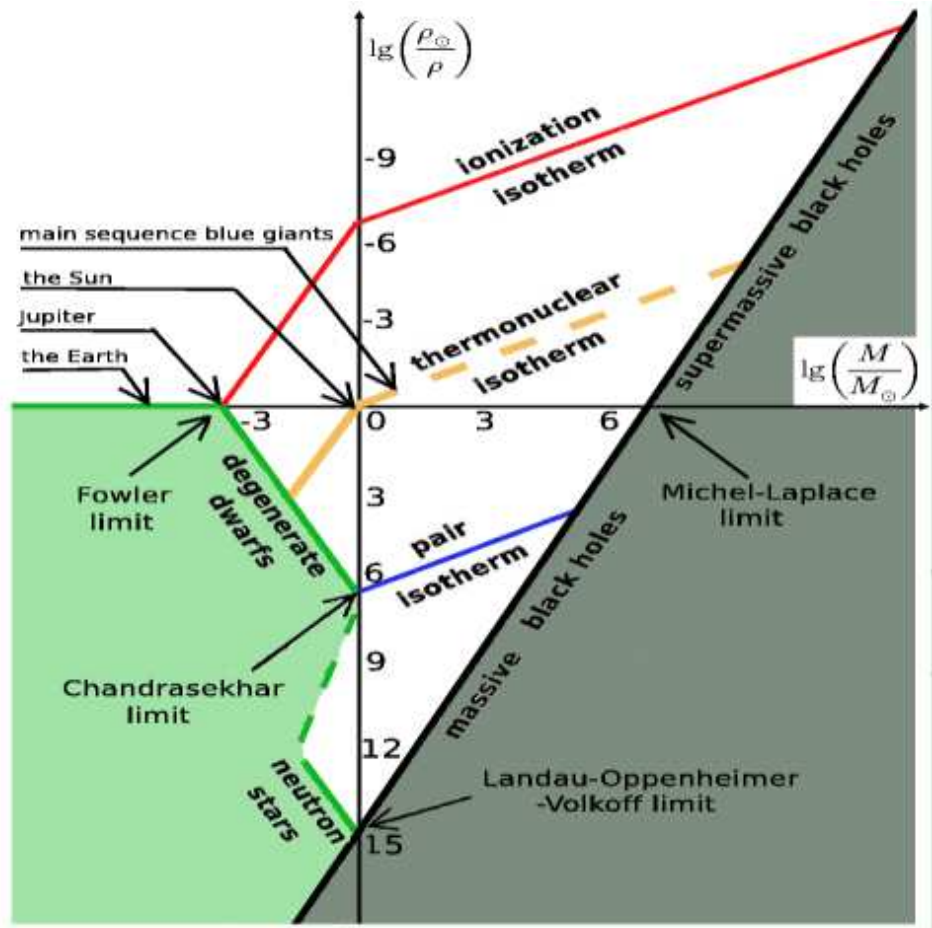, 
height=12.6 cm}
\caption
{
Logarithmic plot of inverse density against mass for the main kinds 
of stellar and planetary bodies, as accounted for \cite{Carter92}
(on the basis of work by pioneers such as Eddington and Chandrasekhar)
in terms just of the masses and charge of the proton and the electron.
}
\label{Fig1}
\end{figure}

To explain Dirac's coincidence on the basis of the anthropic line 
of reasoning that was implicitly followed by Dicke, it suffices to 
work out the typical lifetime $\ttau_\star$ of a main sequence star,  
something that was first understood, on the basis of work by 
Eddington and others, about the same time as Dirac was clarifying the 
essential principles of quantum theory.  Since a star is held 
together by gravity, it is not surprising (see my recent 
recapitulation \cite{Carter92}) that its properties should be 
essentially determined by the gravitational coupling strength: the 
essential conclusions are that typically (give or take one or two 
factors of ten) the mass $\MM_\star$ will be given roughly by 
$\MM_\star\approx \mm_{\rm p}^{-2}$, while its lifetime, which is 
what we are principally concerned with here, will be given by 
$\ttau_\star\approx \varepsilon_{_{\rm N}}\MM_\star/\LL$ where
$\LL$ is its luminosity, and $\varepsilo_{_{\rm N}}$ is the
nuclear binding energy which, on account of  (\ref{nucmarge}), 
will be expressible simply as
$\varepsilo_{_{\rm N}}\approx (2 \mm_\pi/\mm_{\rm p})^2\, .$
The luminosity is sensitive to the stellar mass,
and is given by a rather complicated formula \cite{Carter67}
for small slow burning stars. However for larger brighter
stars of the kind that manufactured our heavy elements
one can use the simple Eddington luminosity formula $\LL/\MM_\star
\approx \mm_{\rm p}\, \mm_{\rm e}^{\, 2}/e^4$ which gives
{\be \ttau_\star\approx (2\, e^2\,\mm_\pi/\mm_{\rm e})^2
\mm_{\rm p}^{- 3}\, .\fe}

Since (in our part of the universe, if not elsewhere) the factor
$ e^2\,\mm_\pi/\mm_{\rm e}$ happens to be of order unity, it is
evident that (at the admittedly rather crude level of accuracy 
involved) the agreement with (\ref{unage}) is perfect: the 
coincidence is explained, and the essential role of gravity confirmed. 
However instead of providing evidence for a radically new theory 
as foreseen by Dirac, the explanation merely confirms what was 
already the established understanding of the situation. Dirac's 
persistent refusal to accept this (from his point of view 
disappointing) outcome shows that, despite the fact that it 
ultimately contributes nothing new in such a case, the  anthropic 
principle is not  merely a tautology. Indeed, whereas Dirac's 
inclination was to spread the a priori probability measure too 
widely, a more common kind of deviation from the anthropic principle 
is to spread it too restrictively. Both kinds of deviation tend 
to be motivated by wishful thinking, typically unwillingness to 
accept the limitations, particularly concerning future prospects, 
that are involved in unbiased application 
of the anthropic principle. 

\section{Anthropic or autocentric forecast}

An extreme, but not unusual, example of a non-anthropic principle 
of the restrictive kind is that of what I would call the autocentric 
principle, whose application consists in a (logically admissible, but 
for predictive purposes sterile) refusal to attribute any a priori 
probability at all to situations other than that in which one has 
already found oneself a posteriori. (As a recent proponent 
\cite{Hartle07} puts it ``what other hypothetical observers with data 
different from ours might see, how many of them there are, and what 
properties they might or might not share with us are irrelevant'')

\begin{figure}
\centering
\epsfig{figure=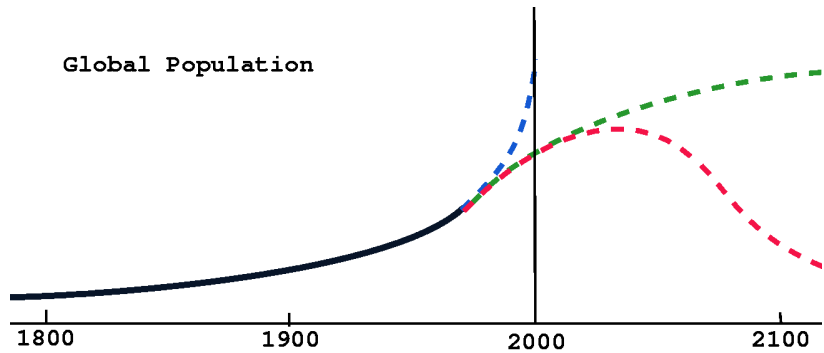, width=10.8 cm}
\caption
{Future of terrestrial human population, as envisaged circa 1980.
The high scenario of continued quasi-exponential growth is already 
excluded by observation, leaving as possibilities the medium scenario 
of sigmoid growth toward saturation at ``sustainable'' level -- 
(which is however unlikely according to anthropic reasoning) and 
the low scenario of quasi-normal evolution (which is most realistic 
in view of the  exhaustion of non-renewable resources).
}
\label{FIG2}
\end{figure}

Recourse to such an autocentric standpoint may be a temptation for 
advocates \cite{Dyson96} of scenarios in which the population of 
our terrestrial civilisation is maintained indefinitely at a
sustainable level (as envisaged by ecologists) or even conceived 
to undergo continued growth (as desired by economists). Whereas 
the autocentric principle is compatible with (neither encouraging nor 
discouraging) such scenarios, on the other hand, according to the 
anthropic principle, the integrated future population of our 
civilisation can not be expected to greatly exceed the number that 
have lived so far, which leaves us with the choice between a sudden 
cut off (what has been called  \cite{Leslie92} the doomsday scenario, 
due to a catastrophe like a global war or some kind of environmental 
disaster) or the less apocalyptic alternative of a gentle and 
controlled decline in population, of the kind that seems already to 
have begun in some developed countries. It is to be noted that since 
I first made such a prediction, about a quarter of a century ago 
\cite{Carter83}, there has already been an inflection, whereby, 
according to official United Nations statistics, see Figure 
\ref{FIG2}, the second time derivative of the global human 
population has changed sign and become negative. It is also to be 
noted that this anthropic prediction concurs with what is to be 
expected anyway on the basis of environmental and other 
considerations, so much so as to be almost redundant, in the sense 
that its conclusion can not be easily escaped even by recourse to 
autocentrism.

\section{Understanding our past evolution}

\begin{figure}
\centering
\epsfig{figure=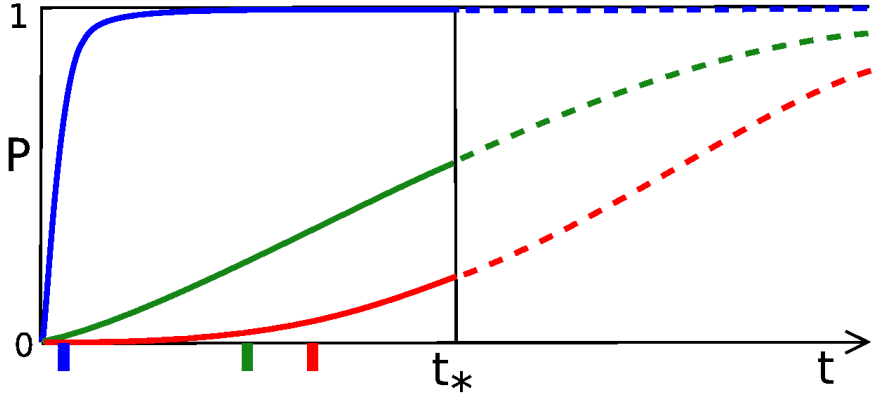, width=12.4 cm}
\caption
{Probability distribution and expectation value for anthropic 
arrival time, firstly if number of difficult steps is zero (high
curve), secondly if it is one (middle curve) and thirdly if it is
two (low curve), using dotted lines for extrapolation beyond the 
cut-off imposed by the Sun's lifetime. The expectation value in 
the first case is far too low to account for the observation that 
the present age of the earth is about half the calculated cut-off 
time. However the second case, that of a single difficult step, 
fits very well, and it is still possible to envisage that there 
may have been two difficult steps (or even more if -- to allow 
for rising of the solar temperature -- the standard cut-off 
calculation needs downward revision).
}
\label{FIG3}
\end{figure}

Although it is perhaps of less immediate practical importance, a more 
intellectually interesting application \cite{Carter83} of the 
anthropic principle is concerned with another remarkable coincidence
concerning the stellar lifetime discussed above, and in particular
with the more precise value that can be given to it when one is 
concerned not with the whole category of main sequence stars but with 
a single specific case, namely that of our own Sun. The remarkable 
coincidence (much more precise than the vague order of magnitude 
relation that fascinated Dirac) is that the predicted total hydrogen 
burning lifetime of the Sun is only about twice the present 
geologically measured age of our planet Earth. What this means is 
that the stochastic biological evolution process whereby our (at one 
stage single celled, then worm like, fish like, and finally 
mammalian) ancestors  developed to become our anthropoid selves took 
fully half the astrophysically available time. If this very 
complicated and technically (unlike stellar astrophysics) not at all 
well understood process had been slower by a modest factor of two our 
civilisation here would never have got to exist at all. It is not at 
all plausible that the intrinsic stochastic mechanism of such a 
biological process should have been tuned a priori so as to agree 
with the externally imposed astrophysical timescale.

What  could have been expected a priori is that in the given 
planetary environment, the stochastically expected time scale, 
$\overline \tt$ say, for  evolution to our anthropically advanced 
state would be either very short or else very long compared with 
the relevant stellar lifetime. In the former case advanced 
civilisations would be of rather common occurrence and the anthropic 
weighting factor would favour their emergence when the relevant star 
was relatively young. Since that is not what we observe in our own 
case, we are left, as the only plausible alternative, with the 
conclusion that the available time is short compared with the 
stochastically expected timescale, $\overline \tt$, which implies 
that (whereas primitive life may be be relatively common 
\cite{Lineweaver02}) advanced civilisations like ours will occur 
only in rare cases for which there was an exceptional run of luck.
More detailed conclusions can be obtained by classifying the 
intermediate evolutionary steps as easy ones, meaning those with a 
high chance of going through in the time available, and difficult 
ones, meaning those with a low chance of going through in the time 
available. Our observation that the Sun is no longer young implies 
that at least one of the steps must have been  of the difficult kind, 
but on the other hand the number of difficult steps can not have 
been very large since if it had the remaining main sequence time  
would have been expected to be much smaller than, not comparable 
with, the time elapsed so far.

\section{Our planetary environment}

The smallest timescales of which we have any actual experimental
or observational knowledge are those of nuclear physical 
processes for which the relevant timescales are of the order of 
$\mm_{\rm p}^{-1}$ meaning about $10^{20}$ in Planck units. The very 
long cosmological, stellar, and biological evolution timescales 
whose anthropic relationships have been discussed are of the order 
of $\mm_{\rm p}^{-3}$ meaning about $10^{60}$ in Planck units. 
Between these very large and very small values, the minimum 
(fraction of a second) perception timescale $\widetilde\ttau$ 
referred to in (\ref{forty}) is given by the geometric mean, 
 namely about $10^{40}$ in Planck units. As this happens to 
be what is recognisably expressible as $\mm_{\rm p}^{-2}$ we 
again find ourselves confronted with the question of whether 
this is just an accidental coincidence, or whether, as with 
Dirac's coincidence it really is explicable in terms of 
gravitational coupling.
 
A significant step toward such an explanation is contained in the 
pioneering investigation of the physical and astrophysical 
limitations on human space dimensions provided \cite{Press80}
by Press, whose key point was that the dimensions of the host planet 
are rather tightly restricted by the requirement that  the 
gravitational field should be strong enough for binding of
water and heavier molecules, but not quite strong enough for
binding of hydrogen. This means that,  compared with the square 
of the escape velocity, the thermal energy factor given for the 
atmospheric nitrogen and oxigen by the square of the sound speed 
has to be rather (but not too) small --  by a factor of 
roughly the order of a thousandth.

For marginal gravitational binding of hydrogen atoms 
(with thermal velocity not far below the escape velocity) 
the generic virial equilibrium formula -- as explained in 
my recent recapitulation \cite{Carter92} --  for a planetary or 
non-relativistic stellar type body of mass $\MM$ density $\rrho$ 
and temperature $\Ttheta$, takes a form given roughly by 
$\MM\approx 10\, \mm_{\rm p}^{-3/2}\, \rrho^{-1/2} \,\Ttheta^{3/2}$. 
Since on a solid or liquid planet the 
typical interparticle separation will be given by the Bohr radius, 
$\ee^{-2} \mm_{\rm e}^{-1}$, the ensuing mass density density will be 
a few times that of water, with order of magnitude 
$\rrho\approx \ee^6 \, \mm_{\rm e}^3 \, \mm_{\rm p}$,  so for
hydrogen binding to be marginal the planetary mass must be given 
roughly by
$ \MM\approx 10\, \mm_{\rm p}^{-2}(\Ttheta/\ee^2 \mm_{\rm e})^{3/2}
\, .$
Now in order for  water to exist in liquid form, the temperature 
must be small, but not extremely small, compared with the Rydberg 
(electronic binding) energy, and
therefor given  by an expression of the form 
{\be \Ttheta_\oplus\approx \frac{_1}{^2}\, \epsilon \, 
\ee^4 \,\mm_{\rm e}\, ,\fe}
 where $\epsilon$ is a numerical factor that was taken by Press to be
about $3\times 10^{-3}$. On the basis of these considerations,
Press obtained for the radius of an earth like planet an expression
of the form
{\be  \RR_\oplus\approx 2 \,\epsilon^{1/2} \,\ee^{-1} \,
\mm_{\rm e}^{-1} \mm_{\rm p}^{-1} \, .\label{earthrad}\fe}
This corresponds to a mass of given by an expression of the form
{\be \MM_\oplus\approx 8\,\epsilon^{3/2}\,\ee^3 \,\mm_{\rm p}^{-2}
\, ,\fe}
which is smaller that the value $\MM_\odot\approx m_{\rm p}^{-2}$ 
that characterises a typical main sequence star such as the Sun by a 
factor of order $\epsilon^{3/2} \ee^3$.

To make the link with local biology what one needs is not the global 
quantities given by the preceding Press formulae, but the value of 
the local Galilean acceleration field given (according to the 
gravitation law discovered by Hooke but subsequently named after 
Newton) by $\gg\approx \MM_\oplus/\RR_\oplus^{\,2}$.  By a convenient 
cancellation (that does not seem to have been previously noticed) it
turns out that the result depends on the mass only of the
electron, not the proton, taking the remarkably simple form
{\be   \gg\approx 2\, \epsilon^{1/2} \, \ee^5 \, \mm_{\rm e}^{\, 2} 
\, ,\label{smallg}\fe}
which is only weakly dependent on the small arithmetical factor
$\epsilon$, but strongly dependent on the electron charge $\ee$.

\section{The weakness of animal flesh.}

Having thus discovered what governs the value of our local 
gravitational field, we are now faced with the less simple question 
of how the resulting value of $\gg$ affects biological organisms 
of the kind to which we belong. This is an issue that I think 
deserves detailed biophysical investigation. Deviating at this point 
from the approach followed by Press \cite{Press80} -- and also from a 
related approach recently developed by Page \cite{Page07} -- my own 
suggestion is that one should think in terms of a basic  biological 
-- or to be more specific, zoological -- characteristic velocity, 
$\widetilde \vv$, given -- as a small fraction of the speed of 
ordinary sound at the relevant temperature, by a relation of the form 
{\be \widetilde \mm\, \widetilde \vv^2\approx \Ttheta_\oplus\, ,\fe}
in which $\widetilde \mm$ is a mass scale characterising relevant 
large biochemical molecules such as proteins. This means that it will
be expressible by a relation of the form
{\be  \widetilde \mm\approx \widetilde\epsilon^{-1} \mm_{\rm p} 
\, ,\label{biom}\fe}
in which (like the Press coefficient $\epsilon$ introduced above)
the quantity $\widetilde\epsilon$ is a small arithmetical factor that
-- for $\widetilde \mm$ to be the mass of a typical protein molecule 
-- should have order of magnitude $\widetilde\epsilon
\approx 0.3\times 10^{-4}$.

This quantity $\widetilde\epsilon$ has the same status as that of
the Press coefficient $\epsilon$, in that the smallness of these
quantities is just an arithmetical (in principle calculable) measure 
of the complexity of the (molecular, not just atomic) systems 
involved, and -- contrary to a notion that was suggested, but 
justifiably criticised by Peierls, in a related discussion 
\cite{PL83} -- it has nothing to do with the smallness of the fine 
structure constant $\ee^2$ (whose role is significant only when heavy
metals are involved) nor of the ratio $\mm_{\rm e}/\mm_{\rm p}$
(which is too small to matter much except for the lightest elements,
namely pure hydrogen and helium, at temperatures far too
low for ordinary life). 

The foregoing estimate for $\widetilde\epsilon$ is actually 
interpretable as meaning that the corresponding zoological 
characteristic velocity, 
{\be \widetilde \vv \approx \ee^2 (\tilde\epsilon\,\epsilon\, 
\mm_{\rm e}/\mm_{\rm p})^{1/2}\, ,\label{biov}\fe}
will in our case be about 3 percent of the speed of ordinary sound. 
This relatively slow speed is mainly attributable to the very small
value of the foregoing estimate for $\widetilde\epsilon \, ,$ which 
has nothing to do with the values of physically adjustable coupling 
constants, but is an ineluctable concomitant of the flabbiness of 
flexible flesh, or even cartilage, as contrasted with the rigidity 
of woody celluloid matter in plants, for which the corresponding 
botanical characteristic velocity would be considerably higher, 
though still subsonic. \,(To emulate the strength of vegetable matter, 
animal bodies do of course incorporate rigid bone structure, but 
the extent to which that is feasible is limited by the ensuing 
sacrifice of flexibility and mobility. I therefore differ from
Press~\cite{Press80} in my opinion that the properties of 
bone itself are of secondary importance, and that the essential
restrictions on animal size are attributable to the finite strength
of the flexible tissues that hold the bones in place.)

Assuming that such a velocity $\widetilde \vv$ (of the order 
of 10 m/sec in our own case) characterises the relevant energies, 
pressures, and tensions (as involved for example in the pumping of 
blood) in an animal body, it will provides a rough upper limit, 
{\be 2\,\gg\, \lel \lta \widetilde \vv^2\, ,\label{vlim}\fe} 
on the supportable value of the gravitational energy per unit 
mass associated with a height difference $\lel$ between different 
parts of the body of the organism. When applied to solid crystalline 
matter, for which the relevant velocity will be of the order of that 
of sound (whose square, as remarked above, must be of the order of 
a thousandth of the square of the escape velocity)
analogous reasoning~\cite{Weisskopf86} indicates that the maximum
possible height of a mountain will be comparable with the thickness of 
the bulk of the atmosphere, and thus of the order of a thousandth
of the planetary radius, which in our terrestrial case provides
an (observationally verifiable) estimate of the  of the order 
of 10$^4$ metres.

	It is instructive to see how the limit (\ref{vlim})
can be derived in a manner similar to that used by Press 
\cite{Press80}, who considered the total energy, $\widetilde\EE$ say, 
needed to break the bonds with energy $\Ttheta_\oplus$ binding the 
molecules on a 2-dimensional shear-disruption surface with area 
$\lel^2$. In terms of the relevant molecular dimension 
$\widetilde\aa$ say (which for the large protein molecules 
considered here will be some tens of times larger than the Bohr 
radius $\ee^{-2}\mm_{\rm e}^{-1}$) the number of molecules in the 
surface layer will be of order $(\lel/\widetilde\aa)^3$ which gives 
$\widetilde\EE\approx \Ttheta_\oplus(\lel/\widetilde\aa)^2\, .$ 
This disruption energy has to be supplied by the action of gravity 
on the mass, $\widetilde\MM$ say, in the corresponding volume of order
$\lel^3\, ,$ which will be given by $\widetilde\MM\approx 
\widetilde\mm(\lel/\widetilde\aa)^3\, .$ The viability condition 
proposed by Press was that the energy liberated in an animal's
fall through its own height $\lel$ should be insufficient to provide
the disruption energy, which gives a limit of the form
$\gg\widetilde\MM\lel\lta\widetilde\EE\, .$

My own opinion is however that animals can learn how to take care
to avoid such dynamical falls, and that what really matters for 
viability is the condition for static support against gravity, which 
will hold so long as the disruption energy cannot be provided by a 
displacement comparable with half the molecular separation distance 
$\widetilde\aa$, which means that instead of the preceding Press 
type inequality one gets a limit of the form
$\gg\widetilde\MM\widetilde\aa/2\lta\widetilde\EE\, ,$ with the 
factor $\lel$ replaced by $\widetilde\aa/2$. This replacement does 
not of course eliminate the dependence on $\lel\, ,$ which is 
implicitly involved through both $\widetilde\EE$ and $\widetilde\MM$. 
Ultimately it is the dependence on $\widetilde\aa$ that cancels 
out, leaving a static support condition of the simple form 
 (\ref{vlim}).

This simple result contrasts with what would be obtained from a 
dynamical survivability condition of the kind proposed by Press, 
which reduces to the not quite so simple form
 $\gg\, \lel^2 \lta \widetilde\aa\,\widetilde \vv^2\, .$
For the actual application \cite{Press80} of this latter formula
to hard bone  -- instead of the fleshy tissue considered 
here -- the molecular radius $\widetilde\aa$ has to be replaced by 
the ordinary Bohr radius $\aa_0=\ee^{-2}\mm_{\rm e}^{-1}$ and 
the very low value of the zoological velocity $\widetilde \vv$ 
used here has to be replaced by the much larger velocity value
that is obtainable from (\ref{biom}) or (\ref{biov}).
simply by setting $\widetilde\epsilon$ to unity. 

\section{Characteristic timescale of human perception}

On the basis of the static support condition (\ref{vlim}),
the preceding considerations imply that (whereas a land plant
or a sea animal may be able to be larger) a land animal will 
be able to have at most a maximum size $\widetilde\lel$ and a 
corresponding biological clock timescale $\widetilde\ttau$ 
given by 
{\be \widetilde\ttau\approx \frac{\widetilde\lel}{\widetilde \vv}
\approx   \frac{\widetilde \vv}{2\gg}\, .\fe}
Using the preceding estimates (\ref{smallg}) and (\ref{biov}) for 
$\gg$ and $\widetilde \vv$ one obtains the formula
{\be \widetilde \ttau \approx 1/\left({4\,e^{3} \, 
\mm_{\rm e}^{3/2}\, \widetilde \mm^{\,1/2}} \right) \, ,\fe}
which is independent of the previously introduced Press coefficient 
$\epsilon$. It does however have  a weak dependence on the  newly 
introduced  coefficient $\widetilde\epsilon$ when expressed in terms 
of the mean coupling constant $\langle \mm\rangle^2\approx 10^{-40}$ 
defined by (\ref{meanm}), taking the form
{\be  \widetilde\ttau \approx \frac{ 
\left(\widetilde \epsilon\, {\mm_{\rm p}/ \mm_{\rm e}}\right)^{1/2}
}{4\, \ee^3 
\langle \mm\rangle^{2}}\, .\label{biotime}\fe}
Since the factor $1/4\,\ee^{3}$ is only of the order of a hundred,
and the dimensionless combination  
$ \widetilde\epsilon\, {\mm_{\rm p}/ \mm_{\rm e}}$ can be expected to 
be rather smaller than unity, it is the factor $1/\langle 
\mm\rangle^{2}$ that is overwhelmingly dominant, so the expectation 
of an essentially gravitational explanation is confirmed.
  
\section{Strong anthropic reasoning}

The idea of what I called the strong anthropic principle
\cite{Carter74} was to extend the arena of application of the weak
anthropic principle to scenarios in which the relevant
(anthropically weighted) a priori probability measure is not 
limited to the part of the universe about which we have direct 
observational knowledge, but extended to other hypothetically 
existing parts of what may be termed a multiverse \cite{Davies04}, 
in which fundamental parameters, such as the fine structure constant 
$\ee^2$ and the gravitational coupling constant
$\langle \mm\rangle^2$,  might have values different from
those (respectively $1/137$ and $10^{-40}$) with which we are 
familiar. In particular I suggested that given the value of 
the former (electromagnetic) coupling constant the weakness 
of the latter (gravitational) coupling constant might be 
explicable in such a framework as due to a selection effect 
-- giving it the maximum value, proportional to a very high 
(the twelfth) power of the fine structure constant -- on the basis 
of the requirement of stellar convection \cite{Carter67} as a 
prerequisite for the necessary planetary formation.

Starting with the application to the marginal nuclear binding
condition (\ref{nucmarge}),
arguments of this strong anthropic kind have since been put 
forward \cite{Carr79,Hogan00,Page03} to account for relations 
involving other parameters, such as those controlling weak 
interactions. However -- in the absence of plausible mechanisms
for the cosmological parameter variations that had to be invoked 
-- such explanations did not become fashionable until the 
situation was revolutionised \cite{Kallosh03} on the theoretical 
side by the rise of modern superstring theories and the attempts 
to unify them in M-theory, and on the observational side by the 
discovery (which surprised everyone, including superstring 
theorists) that the expansion of the universe is not undergoing 
gravitational deceleration but actually accelerating.

The logic leading to the formula (\ref{biotime}) for the 
zoological clock timescale  $\widetilde\ttau$, and to the 
corresponding space dimension, 
{\be \widetilde\lel\approx \epsilon^{1/2}\,\widetilde\epsilon/\left(
4\, \ee\,  \mm_{\rm e}\, \mm_{\rm p} \right)\, ,\label{llength}\fe}
is based on the assumption that, natural selection will tend to 
maximise the latter (within the limits imposed by the ambient 
Galilean gravitational field $\gg$) in order to obtain as large 
as possible a value  for the corresponding corporal particle number, 
as given by $\widetilde \NN\approx\widetilde\nn\,\widetilde\lel^3$ 
in terms of the particle number density $\widetilde \nn\approx 
(\ee^2\, \mm_{\rm e}/2)^3$ of water. The value 
{\be \widetilde\NN\approx (\epsilon^{1/2}\,\widetilde\epsilon\, 
\ee/8\,\mm_{\rm p})^3 \, ,\label{maxno}\fe}
thus obtained for the body  -- of which the brain, in the human case, 
is a significant fraction -- will of course be a wide overestimate of 
what, in view of the haphazard nature of natural selection, is 
actually likely to be achieved in practice, but the limit given by 
(\ref{llength}) has occasionally been approached in a few gigantic 
cases such as that of the brontosaurus. As a fraction of the
number $\NN_\oplus\doteq\MM_\oplus/\mm_{\rm p}$ this is expressible
as the relation
{\be \widetilde\NN/\NN_\oplus\approx (\widetilde\epsilon/16)^3 
\approx 10^{-17}\, ,\fe}
in which the right hand side is a purely arithmetical quantity
whose extremely small value does not depend on any empirical
parameter (such as the proton mass $\mm_{\rm p}$ with which
it is numerically comparable) but is just a consequence of the 
complexity of biochemical processes.  Its cube root -- of the order 
of a millionth -- characterises the maximum size ratio, 
$ \widetilde\lel/\RR_\oplus \, ,$ for a land  animal with the same 
kind of biochemistry as ours, not just on any inhabitable planets 
within our own or nearby galaxies, but even in other parts of the 
multiverse (where quantities such as $\ee$ and $\mm_{\rm p}$ need not
have the values that are familiar). More particularly and memorably,
on any such planet, the maximum land animal size $\widetilde\lel$ 
will be of the order of a thousandth of the atmospheric thickness 
(and the maximum mountain height) which -- as remarked above in the 
discussion of (\ref{vlim}) -- will itself be of the order of a 
thousandth of the planetary radius $\RR_\oplus$ given by 
(\ref{earthrad}).
 
In conclusion, assuming that mental processing benefits from
maximisation of the number of particles in the brain and therefor 
of its host body, it can be seen from (\ref{maxno}) that the strong
anthropic principle will favor regions of the multiverse in which 
the ratio $\ee/\mm_{\rm p}$ is very large, that is to say
where the ratio of gravitational to electric coupling is very small.

\section{Acknowledgments} 

I wish to thank Don Page for many discussions that have
helped me to clarify the various issues dealt with here.

\end{document}